\documentclass[11pt]{article}
\usepackage{amsfonts,amsmath,amssymb}
\usepackage{enumerate}
\usepackage{hyperref}
\usepackage{bbm}
\usepackage{nicefrac}
\usepackage[all]{xy}
\usepackage{simplewick}
\usepackage{graphicx}
\usepackage{bm}
\usepackage{makecell}
\usepackage[table]{xcolor}

\usepackage{upgreek}

\usepackage{booktabs}

\newcommand{\be}{\begin{equation}}
\newcommand{\ee}{\end{equation}}

\newcommand{\bea}{\begin{eqnarray}}
\newcommand{\eea}{\end{eqnarray}}

\newcommand{\bes}{\begin{subequations}}
\newcommand{\ees}{\end{subequations}}

\newcommand{\nn}{\nonumber\\}

\def\sst#1{{\scriptscriptstyle #1}}

\def\0{{\sst{(0)}}}
\def\1{{\sst{(1)}}}
\def\2{{\sst{(2)}}}
\def\3{{\sst{(3)}}}
\def\4{{\sst{(4)}}}
\def\5{{\sst{(5)}}}
\def\6{{\sst{(6)}}}
\def\7{{\sst{(7)}}}
\def\8{{\sst{(8)}}}

\allowdisplaybreaks  

\usepackage{multirow}
\usepackage{rotating}

\newcommand{\ba}{\begin{align}}
\newcommand{\ea}{\end{align}}

\newcommand{\bse}{\begin{subequations}}
\newcommand{\ese}{\end{subequations}}

\allowdisplaybreaks

\begin{document}

\makeatletter
\renewcommand{\theequation}{\thesection.\arabic{equation}}
\@addtoreset{equation}{section}
\makeatother

\begin{titlepage}

\begin{flushright}
\end{flushright}

\vspace{25pt}

   \begin{center}
   \baselineskip=16pt

   \begin{Large}\textbf{
Scalar CFTs and Their Large N Limits}
   \end{Large}

\vspace{25pt}
		
{\large  Junchen Rong$^{1}$ \,and\,  Ning Su$^{2}$}
		
\vspace{30pt}

	\begin{small}
	
	{\it $^{1}$  Fields, Gravity \& Strings, Center for Theoretical Physics of the Universe \\
	Institute for Basic Sciences, Daejeon 305-811, Korea}

	\vspace{15pt}
          
   {\it $^{2}$  CAS Key Laboratory of Theoretical Physics, Institute of Theoretical Physics, \\Chinese Academy of Sciences, Beijing 100190, China}



	\end{small}

\vskip 50pt

\end{center}

\begin{center}
\textbf{Abstract}
\end{center}

\begin{quote}

We study scalar conformal field theories whose large $N$ spectrum is fixed by the operator dimensions of either Ising model or Lee-Yang edge singularity. Using numerical bootstrap to study CFTs with $S_N\otimes Z_2$ symmetry, we find a series of kinks whose locations approach $(\Delta^{\text{Ising}}_{\sigma},\Delta^{\text{Ising}}_{\epsilon})$ at $N\rightarrow \infty$. Setting $N=4$, we study the cubic anisotropic fixed point with three spin components. As byproducts of our numerical bootstrap work, we discover another series of kinks whose identification with previous known CFTs remains a mystery. We also show that ``minimal models'' of $\mathcal{W}_3$ algebra saturate the numerical bootstrap bounds of CFTs with $S_3$ symmetry.

\end{quote}

\vfill

\end{titlepage}

\tableofcontents



\section{Introduction}
Scalar field theories are useful in studying phase transition and critical phenomenon. A large number of these models has been applied to different condense matter systems to extract the critical exponents \cite{Pelissetto:2000ek,Kleinert:2001ax}. The simplest example among them, the $\phi^4$ theory could be used to study phase transition concerning $Z_2$ symmetry breaking, which includes the Ising model \cite{Wilson:1971dc}. 

The critical exponents calculated in field theories are usual based on certain perturbation method, such as $\epsilon$-expansion \cite{Wilson:1971dc} or large $N$ expansion (see \cite{Moshe:2003xn} and references therein). As a non perturbative method, conformal bootstrap program \cite{Ferrara:1973yt,Polyakov:1974gs} has been proven to be useful in studying two dimensional conformal field theories. It has played an important rule in the classification of two dimensional ``minimal models'' \cite{Belavin:1984vu}. In higher dimensions, a significant progress was made in the seminal work of \cite{Rattazzi:2008pe}. There has been a revival of this program after that. An incomplete list of work on conformal bootstrap and related topics is  \cite{Costa:2011mg}. 

Numerical bootstrap is applicable even in regions where neither $\epsilon$-expansion or large $N$ works very well. For three dimensional Ising model, it has provided the most precise the critical exponents so far \cite{ElShowk:2012ht,El-Showk:2014dwa,Kos:2016ysd}. For the perturbative regions, the bootstrap result was also shown to agree with the field theory result. For example, the numerical bootstrap for the scaling dimension of operators in critical $O(N)$ vector model \cite{Kos:2013tga,Kos:2015mba} agrees perfectly with the large $N$ calculation based on the scalar theories \cite{Moshe:2003xn}. The Borel-resummation of  $\epsilon$ expansion series for scaling dimension of operators in critical Ising model also agrees with the bootstrap result \cite{El-Showk:2013nia}.  
 
We will study scalar field theories admitting conformal fixed points whose large $N$ behaviour is controlled by another CFT with central charge of order one. The specific models that we would study are closely related to the continuum limit of the Potts model \cite{Potts:1951rk}. We would like to first consider a scalar theory with quartic interaction in $4-\epsilon$ dimensions. The model was referred as ``restricted Potts model'', and was used as an intermediate steep to study the continuum limit of the Potts model \cite{Zia:1975ha}.   This model was recently revisited in \cite{Osborn:2017ucf}. Its Lagrangian is given by   
\be\label{rescritedPotts}
\mathcal{L}=\frac{1}{2}\partial_{\mu}\phi^i \partial^{\mu}\phi^i+\frac{g_1}{8}d_{ijm}d_{klm}\phi^i\phi^j\phi^k\phi^l+\frac{g_2}{8} (\phi^i\phi^i)^2
\ee
The scalars $\phi^i$ transform in the $n=N-1$ dimensional representation of the symmetric group $S_N$. The totally symmetric tensor $d_{ijk}$ is invariant under the action of $S_N$. The name ``restricted Potts model'' is due to the fact that besides $S_N$, it also preserves an extra $Z_2$ symmetry under which all the scalars change signs. Its symmetry group is therefore slighter bigger than the $S_N$ symmetry of the original Potts model. Suppose one turns on the a trilinear interaction $\frac{1}{3!}d_{ijk}\phi^i\phi^j\phi^k$, the $Z_2$ symmetry is broken and one get the model which describe the continuum limit of the Potts model. The second model that we will consider is a $\phi^3$ theory in $6-2\epsilon$ dimensions, given by 
\be\label{6Dpotts}
\mathcal{L}=\frac{1}{2}\partial_{\mu}\phi^i \partial^{\mu}\phi^i+\frac{g}{6}d_{ijk}\phi^i\phi^j\phi^k
\ee
It could also be used to study the Potts model. As at close to six dimensions, quartic interaction of scalars are irrelevant, and the $\phi^4$ terms in \eqref{rescritedPotts} could be neglected. $N$-state Potts model is known to undergo first order phase transitions for large enough $N$. In accord with this fact, this $\phi^3$ theory is known to have a non-unitary fixed point at imaginary couple $g$. 

The model \eqref{rescritedPotts} is known to have two extra fixed points other than the free theory points and a $O(N)$ invariant point where symmetry got enhanced \cite{Zia:1975ha}. In section \ref{epsilonexpansion}, we look at their operator spectrum to set up the background for later numerical bootstrap study. Taking the large $N$ limit of the $\epsilon$-expansion series for anomalous dimensions and compare with the corresponding series in the Ising model, it can be seen that the scaling dimensions of all the operators that we have studied approach a limit fixed by the scaling dimension of operators in the critical Ising model. The non-unitary fixed point of \eqref{rescritedPotts}, on the other hand, has a large $N$ limit whose operators spectrum is fixed by the Lee-Yang edge singularity. 
 
We then employ numerical bootstrap method to study CFTs with $S_N\otimes Z_2$ global symmetry. We observe that in three dimensions, there indeed exist a series of kinks, whose location at large $N$ approach a point given by the scaling dimension of of the spin operator $\sigma$ and the thermal operator $\epsilon$ in critical Ising model. This confirms the predicted large $N$ behaviour from $\epsilon$-expansion. Setting $N=4$, we were able to observe the famous cubic anisotropic fixed point \cite{AharonyFisher1973, Aharony:1973a,Ketley1973a,Wallace:1974a} with three component spins. Interestingly, the scaling dimension of $\Delta_{\phi}$ agree with with its corresponding value in $O(3)$ invariant Heinsberg model, consisting with the prediction in \cite{Calabrese:2002sz}. As a byproduct of our numerical bootstrap study, we also discover a series of new kinks. We were however not able to identify them with any CFTs with Lagrangian descriptions. By doing numerical bootstrap with $S_3$ symmetry in two dimensions, we have also shown that the ``minimal models'' of $\mathcal{W}_3$ algebra saturates the numerical bootstrap bound. These results are presented in section \ref{bootstrapresult}.

\section{Renormalization of scalar theories}\label{epsilonexpansion}
\subsection{``Restricted Potts Model'' $\rightarrow$ Ising Model }
For the restricted Potts model \eqref{rescritedPotts}, the invariant tensor $d_{ijk}$ could be constructed explicitly according to \cite{Zia:1975ha}, it is possible to define a set of ``vielbeins'' $e_i^{\alpha}$  with $\alpha=1\ldots N$ and $i=1\ldots N-1$ through a recursion relation. These vielbeins tell us how a hyper-tetrahedron with $N$ vertices could be embedded in $N-1$ dimensional space. From group theory point of view, the $N$-dimensional representation is reducible, $N=1\oplus n$. Take $N=3$ as an example, the three vielbeins 
\be
e^1=(\frac{\sqrt{3}}{2},\frac{1}{2}),\quad 
e^2=(-\frac{\sqrt{3}}{2},\frac{1}{2}),\quad 
e^3=(0,1).
\ee
form a equilateral triangle, the symmetric group $S_3$ consists of all $O(2)$ rotation that keeps this triangle invariant. Using $e_i^{\alpha}$, the totally symmetry tensor could be defined as
\bea\label{dtensor}
d_{ijk}&=&\sum_{\alpha}e_i^{\alpha}e_j^{\alpha}e_k^{\alpha},
\eea
The details of the two loop calculation of \eqref{rescritedPotts} is summarised in Appendix \ref{renormalization}, which is based on the general formula in \cite{Fei:2016sgs}. It is in principle easy to extend the result to three loop using the result of \cite{Osborn:2017ucf}. We will however only focus on the two loop results. 

The beta function of this model have in total four fixed points 
\bea\label{fixedpointsPotts}
\text{free theory}:&&\quad g_1=g_2=0,\nn
\text{critical O(n) point}:&&\quad g_1=0,g_2\neq 0,\nn
P_1:&& \quad g_1\neq 0,g_2\neq 0,\nn
P_2:&& \quad g_1\neq 0,g_2\neq 0,
\eea
Since $S_N\otimes Z_2$ is a subgroup of O(n), with $n=N-1$. The O(n) invariant fixed point is also present. We will focus on the two extra new fixed points $P_1$ and $P_2$. The scaling dimensions of the operators we have studied are given in Table \ref{P1dimension} and Table \ref{P2dimension}. The quadratic operators falls into various irreps of the symmetry group $S_N$ (they are clearly $Z_2$ odd), as
\be\label{tensorirrpes}
\text{n}\otimes \text{n} \rightarrow \text{S} \oplus \text{A} \oplus \text{n}\oplus \text{T}'.  
\ee
The irreducible representation n appears as a result of the existence of invariant tensor $d_{ijk}$. It is interesting to observed that for both of the fixed points, the scaling dimensions of low lying operators at the large $N$ limit could be expressed in terms of the Ising model spectrum.
\begin{table}[h]
\centering
\begin{tabular}{|l|l|l|l|}
\hline
Operator      &$Z_2$&$\Delta$ & $\Delta_{n\rightarrow \infty}$                    \\ \hline
$\phi\in $ n        & $-$ &  \eqref{SDphi}      & $\Delta^{\text{Ising}}_{\sigma}$                  \\ \hline
$\phi^2\in $ S &$+$ &   \eqref{SDphi2S}      & $D-\Delta^{\text{Ising}}_{\epsilon}$            \\ \hline
$\phi^4\in $ S, 1st &  +&  \eqref{SDphi4SP1}      & $2\times(D-\Delta^{\text{Ising}}_{\epsilon}$) \\ \hline
$\phi^4\in $ S, 2st & + &  \eqref{SDphi4SP1}        & $\Delta^{\text{Ising}}_{\epsilon'}$               \\ \hline
$\phi^2\in $ n &       +&\eqref{SDphi2n}    & $\Delta^{\text{Ising}}_{\epsilon}$                \\ \hline
$\phi^2\in $ T$'$ &        +& \eqref{SDphi2T}   & $2\times \Delta^{\text{Ising}}_{\sigma}$          \\ \hline
\end{tabular}
\caption{Scaling dimensions of low lying operators at the fixed point $P_1$.}
\label{P1dimension}
\end{table}
\begin{table}[h]
\centering
\begin{tabular}{|l|l|l|l|}
\hline
Operator      &$Z_2$ &$\Delta$ & $\Delta_{n\rightarrow \infty}$                    \\ \hline
$\phi\in $ n        &  $-$  &  \eqref{SDphi}     & $\Delta^{\text{Ising}}_{\sigma}$                  \\ \hline
$\phi^2\in $ S & +& \eqref{SDphi2S}         & $\Delta^{\text{Ising}}_{\epsilon}$            \\ \hline
$\phi^4 \in $ S, 1st &+&   \eqref{SDphi4SP2}         & $2 \times \Delta^{\text{Ising}}_{\epsilon}\quad\quad $\space\space\space  \\ \hline
$\phi^4 \in $ S, 2st & +&  \eqref{SDphi4SP2}         & $\Delta^{\text{Ising}}_{\epsilon'}$               \\ \hline
$\phi^2\in $ n & +& \eqref{SDphi2n}         & $\Delta^{\text{Ising}}_{\epsilon}$                \\ \hline
$\phi^2\in $ T$'$ & +& \eqref{SDphi2T}         & $2\times \Delta^{\text{Ising}}_{\sigma}$          \\ \hline
\end{tabular}
\caption{Scaling dimensions of low lying operators at the fixed point $P_2$.}
\label{P2dimension}
\end{table}

\subsection{The spectrum of (de)coupled CFTs}\label{decoupleCFT}
We should mention that the large $N$ behaviour could already be partially inferred from combining the result of \cite{Osborn:2017ucf} and much earlier work of \cite{aharony1973critical,Emery:1975zz} on cubic anisotropic systems. We will explain this point in the present section, and try to better understand the large $N$ limit. 

In \cite{Osborn:2017ucf}, another $\phi^4$ theory was studied, the model was obtained by replacing the $d_{ijm}d_{klm}$ in \eqref{rescritedPotts} with 
\be\label{NISing}
Q_{ijkl}=\bigg\{ \begin{array}{ll}
                  1,&\quad  \text{if }i=j=k=l, \\
                  0, &\quad \text{otherwise. } 
                \end{array}
\ee
The model  has a long history of being studied \cite{AharonyFisher1973, Aharony:1973a,Ketley1973a,Wallace:1974a,Toledano:a,KLEINERT1995284,Fei:2015oha}, and certain critical exponents are known up to six loops \cite{Carmona:1999rm}. This model preserves a symmetry group which is the generalized symmetric group $S(2,N)=S_N \otimes Z_2^N$.  Like \eqref{rescritedPotts}, it has also four fixed points 
\bea
\text{free theory}:&&\quad g_1=g_2=0,\nn
\text{critical O(N) point} :&&\quad g_1= 0,g_2\neq 0,\nn
\text{cubic anisotropic point}:&& \quad g_1\neq 0,g_2\neq 0,\nn
N\text{ copies of decoupled Ising models}:&& \quad g_1\neq 0,g_2= 0.
\eea
It was shown in \cite{Osborn:2017ucf} that certain numbers that appear in the renormalization calculation of both models have the same large $N$ limit (see Section 5.1.2), and therefore the two models approach the same limt at $N \rightarrow \infty$. The fixed point $P_2$ approaches $N$ copies of decoupled Ising models, it is therefore not surprising that it spectrum are given by the scaling dimensions of operators in the Ising model. 

 It is straightforward to work out the spectrum of the decoupled CFTs. Suppose a certain CFT preserves symmetry group G, then $N$ decoupled copy of this CFT preserves the symmetry group $G\wr S_N=S_N\otimes G^N$. The symbol ``$\wr$'' stands for wreath product, which can be viewed as a short hand notation of the symmetry group. The group $G^N$ acts independently on each copy of the CFTs, while $S_N$ interchange them. We will consider only operators which are invariant under the full group $G\wr S_N$. Suppose the component CFT has the following conformal primaries operators which are invariant under the action of $G$,  
\be
O_1, O_2, O_3, \ldots, 
\ee
The decoupled model then has the following operators which are also invariant under $S_N$ permutations, 
\bea
\mathcal{O}_1=\frac{1}{\sqrt{N}}\sum_{i} O^i_1, \quad \mathcal{O}_2=\frac{1}{\sqrt{N}}\sum_{i} O^i_2, \quad \mathcal{O}_3= \frac{1}{\sqrt{N}}\sum_{i} O^i_3,\quad \ldots,
\eea
where space-times indices are supressed for simplicity. The indices $i$ denote which copy of the CFTs does $O^i$ belong to. Picking two operators from same CFT copy, take $O_1$ and $O_2$ as an example, one could easily make $S_N$ invariant operators of the following form 
\be
\frac{1}{\sqrt{N^2-N}} \sum_{i\neq j} O_1^i O_2^j.
\ee
The coefficient in front of the operators is due to normalization. For operators with spin, the space time indices need to be arrange properly for them to have definite spin. The condition $i\neq j$ makes sure that  the composite operator is made of two operators from different copies of CFTs, so that it  would not be renormalised. The summation over $i\neq j$ pairs makes it $S_N$ invariant. If $O_1$ and $O_2$ are scalars, we could also construct the following operators 
\bea
&&\left[\mathcal{O}_1\mathcal{O}_2\right]_{n=0,l=1}=\frac{1}{\sqrt{N^2-N}}\sum_{i\neq j} \Delta_2(\partial_{\mu} O^i_1) O^j_2- \Delta_1O^j_1(\partial_{\mu} O^i_2) .\nn
&&\left[\mathcal{O}_1\mathcal{O}_2\right]_{n=2,l=0}=\frac{1}{\sqrt{N^2-N}}\sum_{i\neq j} \bigg( \frac{\Delta_1}{2\Delta_1+2-D} (\partial^2 O^i_1) O^j_2\nn&&
\quad\quad\quad\quad\quad\quad\quad\quad\quad\quad\quad \quad\quad\quad   -\partial_{\mu}O^i_1\partial^{\mu}O^j_2 +\frac{\Delta_2}{2\Delta_2+2-D} O^i_1  (\partial^2 O^j_2)\bigg).\nn
&&\ldots.
\eea
We have borrowed the notation $[\mathcal{O}_1\mathcal{O}_2]_{n,l}$ for double trace operator in AdS/CFT context \cite{Heemskerk:2009pn,Heemskerk:2010ty,Fitzpatrick:2011dm,ElShowk:2011ag}.  The scaling dimensions of these operators are simply $\Delta=\Delta_1+\Delta_2+n+l$. The derivatives acting on the operators are arranged so as to ensure that they are conformal primaries. The procedure of choosing an appropriate derivatives structure is exactly the same as constructing conformal primaries for ``generalized free fields'', as studied in  \cite{Fitzpatrick:2011dm}.  One could also follow it to constructed ``double trace'' conformal primaries operators with higher spin and twist. Even though we are not aware of it appearing anywhere in the literature, a similar procedure should exist for constructing double trace operators made of operators with non-zero spins.  It is also interesting to look at the 4-pt function consist of identical scalar operators,
\bea\label{4ptfunction}
&&\langle \mathcal{O}(x_1)\mathcal{O}(x_2)\mathcal{O}(x_3)\mathcal{O}(x_4)\rangle\nn
&&\qquad\qquad=\frac{1}{N^2} \sum_{i,j,k,l} \langle O^i(x_1)O^j(x_2)O^k(x_3)O^l(x_4)\rangle\nn
&&\qquad\qquad=\frac{1}{N^2} \sum_{i=j\neq k=l} \frac{1}{x_{12}^{2\Delta_O}x_{34}^{2\Delta_O}}+\frac{1}{N^2} \sum_{i=k\neq j=l} \frac{1}{x_{13}^{2\Delta_O}x_{24}^{2\Delta_O}}+\frac{1}{N^2} \sum_{i=l\neq j=l}\frac{1}{x_{14}^{2\Delta_O}x_{23}^{2\Delta_O}}\nn
&&\qquad\qquad\quad +\frac{1}{N^2} \sum_{i=j=k=l} \langle O O O O \rangle \nn
&&\qquad\qquad=(1-\frac{1}{N})\bigg(\frac{1}{x_{12}^{2\Delta_O}x_{34}^{2\Delta_O}}+\frac{1}{x_{13}^{2\Delta_O}x_{24}^{2\Delta_O}}+\frac{1}{x_{14}^{2\Delta_O}x_{23}^{2\Delta_O}}\bigg)+\frac{1}{N}\langle  O O O O\rangle.
\eea
The condition $i=j\neq k=l$ in the second line makes sure that $O^i$ and $O^j$ come form the same copy of CFT, while $O^k$ and $O^l$ comes from a different copy. Its contribution to the four point function therefore reduce to to two point functions.
The leading term in $\frac{1}{N}$ expansion clearly factorises into disconnected two point functions, which are the four point function of ``generalized free fields''. It is equivalent to the dual boundary four point function of a free massive scalar with AdS mass $m^2L^2=-\Delta_{\epsilon}^{\text{Ising}}(D-\Delta_{\epsilon}^{\text{Ising}})$ \cite{Heemskerk:2009pn,Fitzpatrick:2011dm}. The sub-leading behaviour receives contribution from both a disconnected piece and a connected piece which are given by the four point function of the component the CFT, as denoted by $\langle  O O O O\rangle$. 

Specialising to the Ising model, the first three operators with spin-0 and lowest scaling dimensions are 
\bea
&\frac{1}{\sqrt{N}}\sum_{i}\epsilon^i,& \qquad \Delta=\Delta^{\text{Ising}}_{\epsilon}\nn
&\frac{1}{\sqrt{2N^2-2N}}\sum_{i\neq j} \epsilon^i \epsilon^j,& \qquad \Delta=2\times \Delta^{\text{Ising}}_{\epsilon}\nn
&\frac{1}{\sqrt{N}}\sum_{i}\epsilon'^i,& \qquad \Delta=\Delta^{\text{Ising}}_{\epsilon'}.
\eea
They have the same scaling dimension as the S-channel operators\footnote{``S-channel operators'' is short for operators transforming in singlet representation of $S_N$} at fixed point $P_2$. See table \ref{P2dimension}.

At the cubic anisotropic fixed point of \eqref{NISing}, the coupling constants become \cite{Aharony:1973a},
\be
g_1= g^{\text{Ising}}+\mathcal{O}(\frac{1}{N}),\quad  g_2=\mathcal{O}(\frac{1}{N}).
\ee 
The action of the model  becomes $N$ copies of Ising model actions plus certain $\mathcal{O}(\frac{1}{N})$ corrections. It can be shown that this is also true for the fixed point $P_1$ of model \eqref{rescritedPotts}. We will sometime refer these large $N$ CFTs as coupled Ising models for obvious reasons.  At large $N$, the renormalization is clearly dominated  by the Ising model coupling, which explains why the scaling dimensions of Ising model operators appear in the spectrum. These fixed points fit into the class of models studied by Victor Emery in \cite{Emery:1975zz}. Their critical exponents are related to Ising critical exponents by \cite{Emery:1975zz,Aharony:1973a,Fisher:1968zzb} 
\be
\eta=\eta^\text{Ising}+ \mathcal{O}(\frac{1}{N}),\quad  \nu=\frac{\nu^\text{Ising}}{1-\alpha^\text{Ising}}+\mathcal{O}(\frac{1}{N}), \quad \text{and}\quad  \alpha=\frac{\alpha^\text{Ising}}{1-\alpha^\text{Ising}}++\mathcal{O}(\frac{1}{N}). 
\ee
Translated into operator dimensions, this means 
\be
\Delta_{\phi} \rightarrow \Delta^{\text{Ising}}_{\sigma}, \quad \text{and} \quad \Delta_{\phi^2\in S}\rightarrow D-\Delta^{\text{Ising}}_{\epsilon}, 
\ee
agreeing exactly with the Table \ref{P1dimension}. What's more, operators like 
\be
\frac{1}{N}\sum_{i} \epsilon^i
\ee  
self average as in the critical O($N$) vector model \cite{Moshe:2003xn}. Its four point function are expected to factorise at large-$N$ limit as in \eqref{4ptfunction}. The spectrum of S-channel operators should be exactly the same as the decoupled Ising point, and also fall into the categories of ``single trace operators'', ``double trace operators'' and so on. The only modification that one need to make is the replacement 
\be\label{changeboundary}
\Delta^{\text{Ising}}_{\epsilon} \rightarrow D-\Delta^{\text{Ising}}_{\epsilon}
\ee
This is again supported by the calculation in Table \ref{P1dimension}, where  an operator with $D-\Delta^{\text{Ising}}_{\epsilon}$ is found to be accompanied by a ``double trace'' operator with the scaling dimension $2\times (D-\Delta^{\text{Ising}}_{\epsilon})$. 
\subsection{Potts Model $\rightarrow$ Lee-Yang Singularity} 
Before closing this section, we briefly mention the large $N$ behaviour of the scalar model \eqref{6Dpotts}, the continuum limit of $N$-state Potts models. The theory has a non-unitary fixed point at generic $N$. It was pointed out in \cite{fortuin1972random} that the $N=1$ limit of $N$-state Potts models gives the percolation model. Therefore people have been using \eqref{6Dpotts} to calculate the critical exponents of the percolation problem \cite{deAlcantaraBonfim:1981sy, Gracey:2015tta}. The three loop renormalization for operator dimensions is summarised in Table \ref{Pottsdimension}. (See Appendix \ref{general3loop6D} for more details.)
\begin{table}[h]
\centering
\begin{tabular}{|l|l|l|}
\hline
Operator      & $\Delta$ & $\Delta_{n\rightarrow \infty}$                    \\ \hline
$\phi\in n$        &    \eqref{phi3all}      & $\Delta^{\text{Lee-Yang}}_{\phi}$                  \\ \hline
$\phi^2\in S$ &    \eqref{phi3all}      & $D-\Delta^{\text{Lee-Yang}}_{\phi}$            \\ \hline
$\phi^3\in S$ &    \eqref{phi3all}      & $\Delta^{\text{Lee-Yang}}_{\phi^3}$ \\ \hline
\end{tabular}
\caption{Scaling dimensions for continuum N-state Potts from $\phi^3$ theory.}
\label{Pottsdimension}
\end{table}
By taking the $N \rightarrow \infty$ limit, it is clear that the scaling dimensions of operators is fixed by the spectrum of Lee-Yang edge singularity CFT.  It can also be shown that the coupling constant at large $N$ is given by 
\be
g=g^{\text{Ising}}+\mathcal{O}(\frac{1}{N}).
\ee
By the same argument as in previous section, operators that are invariant under $S_N$ should fall into the categories of ``single trace operators'', ``double trace operators'' and so on. The single trace spectrum is given by the spectrum of Lee-Yang edge singularity, with the replacement 
\be
\Delta^{\text{Lee-Yang}}_{\phi}\rightarrow D-\Delta^{\text{Lee-Yang}}_{\phi}.
\ee
The operator next to the ones listed in Table \ref{Pottsdimension} should have scaling dimension $2\times(D-\Delta^{\text{Lee-Yang}}_{\phi})$.   

\section{Numerical bootstrap for CFTs with $S_N$ symmetry}\label{bootstrapresult}
\subsection{The fixed point $P_1$ from numerical bootstrap}\label{SNboot}
In this section, we will show that the fixed point $P_1$ studied in previous section could be observed in numerical bootstrap. Conformal bootstrap is based on the crossing symmetry and unitarity. Crossing symmetry means that the following two ways of computing its four point functions should lead to equivalent result
\be\label{crossingsym}
\langle
\contraction{}{\phi_i(x_1)}{}{\phi_j(x_2)}
\phi_i(x_1)\phi_j(x_2)
\contraction{}{\phi_k(x_3)}{}{\phi_{l}(x_4)}
\phi_k(x_3)\phi_{l}(x_4)
\rangle=\langle
\contraction{}{\phi_i(x_1)}{\phi_j(x_2)\phi_k(x_3)}{\phi_{l}(x_4)}
\contraction[2ex]{\phi_i(x_1)}{\phi_i(x_2)}{}{\phi_{l}(x_3)}
\phi_i(x_1)\phi_j(x_2)\phi_k(x_3)\phi_{l}(x_4)
\rangle.
\ee
The lines connecting the operators denote how operator product expansion (OPE) is performed. This is true for any conformal field theories. Unitarity on the other hand requires all the OPE coefficients $\lambda_{O_1 O_2 O_3}$ to be real. 

\begin{figure}[h]
\centering
\includegraphics[scale=0.5]{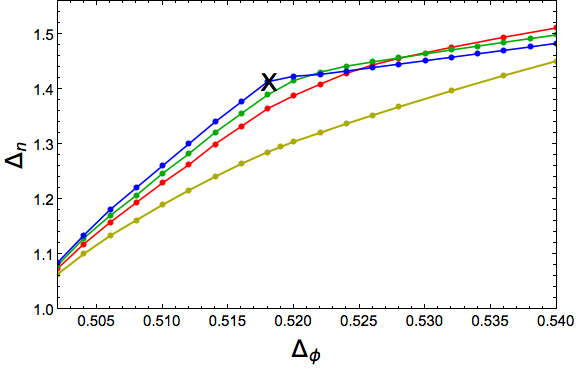}
\caption{Numerical bootstrap bound on the scaling dimension of the first n-channel scalar operators in CFTs with $S_{N}\otimes Z_2$ symmetry (small $\Delta_{\phi}$ region). Yellow, red, green and blue curves are for $N=4,6,10,100$ respectively. The black cross denotes the scaling dimension of Ising model operators $(\Delta^{\text{Ising}}_{\sigma},\Delta^{\text{Ising}}_{\epsilon})$. The bounds is obtained at $\Lambda=19$.}
\label{Snleft}
\end{figure}

By assuming certain conditions on the spectrum of operators that appears in the  OPE 
\be
\phi^i\times \phi^j\sim \sum_{O} O,
\ee
and test the positivity of $\lambda_{\phi\phi O}^2$, one could then check whether such an assumption is consistent with unitarity and crossing symmetry. We will leave the details of how this method was implemented in Appendix \ref{Snsection}. The conditions that we have assumed for the spectrum are:
 \begin{itemize}
 \item{the external operator $\phi^i$ has scaling dimension $\Delta_{\phi}$,}
 \item{the first spin-$0$ operator in the $n$-channel has scaling dimension greater than or equal to $\Delta_{n}$,}
  \item{all the other operators that appear in $\phi^i\times \phi^j$ has scaling dimensions greater than or equal to the unitarity bound.}
 \end{itemize}
We have scanned a certain region of the $(\Delta_{\phi},\Delta_{n})$ plane and the result is presented in Figure \ref{Snleft}. The result is obtained by setting $\Lambda=19$, with the range of spin chosen to be $l\in\{1,\ldots 25\}\cup\{49,50\}$. The region above the curves are excluded, which means there is no unitary CFTs with the assumed spectrum. 
For large enough $N$, a clear kink could be observed in the numerical bootstrap curve. The appearance of kinks in numerical bootstrap is a strong indication of the existence of a conformal field theory. More interestingly, as $N$ increase, the location of the kink approaches the point $(\Delta_{\sigma}^{\text{Ising}},\Delta_{\epsilon}^{\text{Ising}})$, as denoted by the black cross in Figure \ref{Snleft}. This confirms the prediction from from previous section. 

From Table \ref{P1dimension} and \ref{P2dimension}, it is clear that $(\Delta_{\phi}, \Delta_{n})$ should approach $(\Delta_{\sigma}^{\text{Ising}},\Delta_{\epsilon}^{\text{Ising}})$  for both fixed points $P_1$ and $P_2$. We therefore need to determine which one of them corresponds to the kink in Figure \ref{Snleft}. This could be achieved by introducing one extra condition in the assumed spectrum
\begin{itemize}
 \item{The first spin-$0$ operator in the $S$-channel has scaling dimension greater than or equal to $\Delta_{n}+0.1$.}
 \end{itemize}
At large enough $N$, this assumption would clearly exclude point $P_2$, while preserves $P_1$, remember $D-\Delta_{\epsilon}^{\text{Ising}}\approx 1.5874$, while $\Delta_{\epsilon}^{\text{Ising}}\approx 1.4126$. We have checked that the $S_{100}$ curve does not change after introducing this condition, therefore proves that the kink corresponds to fixed point $P_1$.

\begin{figure}[h] 
\centering
\includegraphics[scale=0.5]{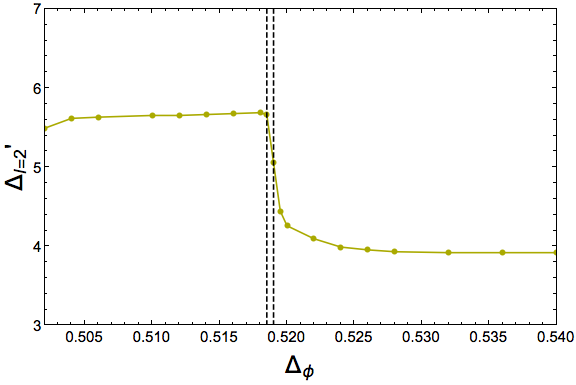}
\caption{Numerical Bootstrap bound on the scaling dimension of the second S-channel spin-2 operator in CFTs with $S_{4}\otimes Z_2$ symmetry. The gap for n-channel scalars has been set to be slightly lower than the maximally allowed value in Figure \ref{Snleft}.  The curve is obtained at $\Lambda=19$. The dashed lines are the estimation of three dimensional O(3) Heisenberg model using Monte Carlo method \cite{Campostrini:2002ky}.}
\label{Snspin2}
\end{figure}

The $N=4$ case deserves some special attention, the symmetry group $S_4\otimes Z_2$ is isomorphic to $S_3\wr Z_2=S_3\otimes Z_2^3$ \cite{Zia:1975ha}. The two group clearly has the same order as $4!\times 2= 3!\times 2^3= 48$. This means that the ``restricted Potts model'' \eqref{rescritedPotts} with $N=4$ is equivalent to the cubic anisotropic model \eqref{NISing} at $N=3$. From Figure \ref{Snleft} itself, it is not clear whether there is a CFT saturating the bootstrap bound or not, since there is no clear kink on the $N=4$ curve. One could study this case more carefully by changing the assumptions of the spectrum into
 \begin{itemize}
 \item{The external operator $\phi^i$ has scaling dimension $\Delta_{\phi}$,}
 \item{the first spin-$0$ operator in the $n$-channel has scaling dimension greater than or equal to $\Delta_{n}=\Delta_{n}^{\text{Max}}-0.002$,}
  \item{the second spin-$2$ operator in the $S$-channel has scaling dimension greater than or equal to $\Delta_{S,l=2}'$ (note the first spin-2 operator needs to be the energy momentum tensor),}
  \item{all other operators that appear in $\phi^i\times \phi^j$ OPE have scaling dimensions greater than or equal to the unitarity bound.}
 \end{itemize}
Notice $\Delta_{n}$ is chosen to be sightly below the maximal allow bound from Figure \ref{Snleft}. This method was introduced in \cite{ElShowk:2012ht} to study the scaling dimension of operator $\epsilon'$ in critical Ising model (see Figure 6). We could similarly carve out the allowed region of $(\Delta_{\phi},\Delta_{S,l=2}') $. This is presented in Figure \ref{Snspin2}.
The dashed lines are the scaling dimension of $\phi^i$ in $O(3)$ invariant Heisenberg model from Monte Carlo simulation \cite{Campostrini:2002ky}. The reason that we could compare the scaling dimension of operators in cubic anisotropic model with operators in $O(3)$ invariant Heisenberg model is that an analysis of the six loop calculation of both models shows that there are surprising cancellations in the different between their critical exponents \cite{Calabrese:2002sz},
\be
\eta^{\text{Cubic}}-\eta^{\text{Heisenberg}}=-0.0001(1),\quad \nu^{\text{Cubic}}-\nu^{\text{Heisenberg}}=-0.0003(3).
\ee
$\Delta_{\phi}$ for cubic anisotropic critical point should be equal to its value for $O(3)$ invariant Heisenberg model to high precision. 
One could clearly observe a sharp drop of the curve at the location of the dashed lines. This curve resembles the bounds for $\Delta_{\epsilon'}$ obtained in \cite{ElShowk:2012ht} for Ising model. We therefore conclude that the cubic fixed point is located at around $\Delta_{\phi}\approx 0.5187$ and saturate the numerical bootstrap bound in Figure \ref{Snleft}.

A long standing question concerning cubic fixed point and the Heinsberg fixed point is the relative height between the two of them along renormalization group flow. This is important experimentally, since the IR CFT is the one that governs the phase transitions. Because of the cancelation mentioned already, it is very hard to distinguish the two CFTs by measuring either $\eta$ or $\nu$. However, the critical exponent corresponding to $\Delta_{n}$, if could be measured, is probably a good candidate. Notice in our case $\Delta_{n}\approx 1.292$, while at the Heinsberg point, $\Delta_{T} \leq 1.22$ is required by numerical bootstrap \cite{Kos:2015mba}. 
\subsection{Other bootstrap results: unidentified kinks}
\begin{figure}[h]
\centering
\includegraphics[scale=0.5]{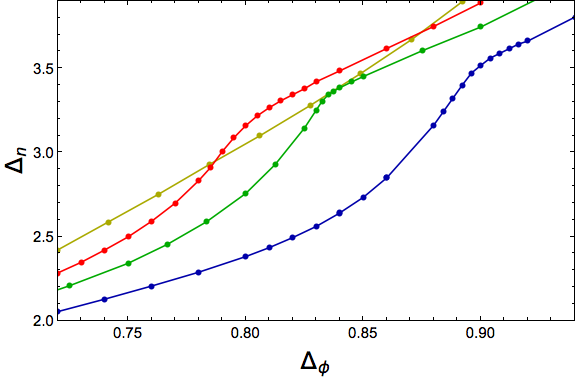}
\caption{Numerical bootstrap bound on the scaling dimension of the first n-channel scalar operators in CFTs with $S_{N}\otimes Z_2$ symmetry (large $\Delta_{\phi}$ region). Yellow, red, green and blue curves are for $N=4,6,10,100$ respectively. The curves are obtained at $\Lambda=23$.}
\label{Snright}
\end{figure}

\begin{figure}[h]
\centering
\includegraphics[scale=0.5]{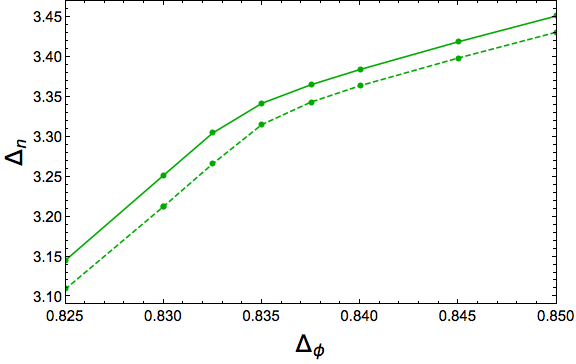}
\caption{Numerical bootstrap bound on the scaling dimension of the first n-channel scalar operators in CFTs with $S_{10}\otimes Z_2$ symmetry. The solid line corresponds to bounds without spin-2 gap, while the dashed shows the result when a small gap for spin-$2$ operator in the $S$-channel is introduced.}
\label{Snrightspin2}
\end{figure}

The study in previous section was focused on the region where $\Delta_{\phi}$ is close to the unitarity bound, it is straight forward to extent the result to the region with much higher $\Delta_{\phi}$. This is presented in Figure \ref{Snright}.
Surprisingly, for large enough $N$, we could again observe some kinks in the numerical bootstrap curve. Unlike those CFTs in previous section, we are not able to find some Lagrangian description for them. Instead, we will show that these kinks pass some consistency checks for them to actually be CFTs. For any full-fledge conformal field theories, it necessarily contains energy momentum tensor in its spectrum. There should be a spin-2 operator saturating the unitarity bound. If the kinks we observed correspond to actual CFTs, they should not survive when a gap is introduced for the spin-2 operators in S-channel.
This fact is tested by considering adding the following condition in the assumed spectrum
 \begin{itemize}
  \item{the first spin-$2$ operator in the $S$-channel has scaling dimension greater than or equal to $3.05$,}
 \end{itemize} 
Taking the $N=10$ curve as an examples, the allowed region for $(\Delta_{\phi},\Delta_{n})$ is presented in Figure \ref{Snrightspin2}. The solid line corresponds to the result without the above condition, while for the dashed line, above condition is included. Clearly, when the gap for spin-2 operator is imposed, the curve moves downward, showing that energy momentum tensor is present in the spectrum. 

\subsection{Other bootstrap results: ``minimal models'' of $\mathcal{W}_3$ algebra}\label{W3result}
The crossing equations we derived in Appendix \ref{Snsection} apply to CFTs with $S_N\otimes Z_2$ symmetry. It could also be easily generalized to study CFTs with $S_N$ symmetry, this is simply achieved by change the assumed spectrum to be  
 \begin{itemize}
 \item{The external operator $\phi^i$ has scaling dimension $\Delta_{\phi}$,}
 \item{the first spin-$0$ operator in the $n$-channel has scaling dimension $\Delta_{\phi}$, while the second spin-$0$ operator in the $n$-channel has scaling dimension greater than or equal to $\Delta_{n}'$,}
  \item{all other operators that appear in $\phi^i\times \phi^j$ has scaling dimensions greater than or equal to the unitarity bound.}
 \end{itemize}
Notice since $d_{ijm}$ is an invariant tensor of $S_N$ group (which is not invariant under $S_N \otimes Z_2$), scalar operator $\phi^i$ would appear in its own OPE, $\phi^i \times \phi^j\sim d_{ijk} \phi^k$. 

\begin{figure}[h]
\centering
\includegraphics[scale=0.5]{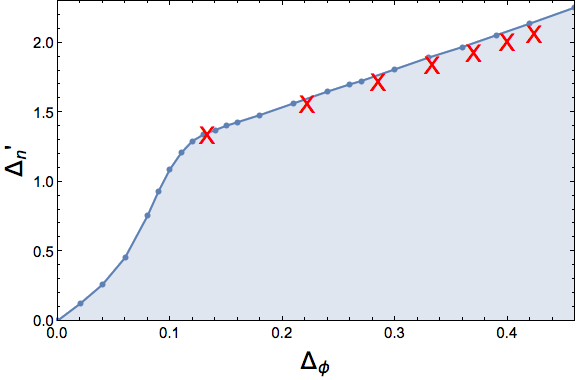}
\caption{Numerical bootstrap bound on the scaling dimension of the second n-channel scalar operators in CFTs with $S_3$ symmetry. The crosses correspond to minimal models with $\mathcal{W}_3$ algebra. The first cross to the left is 3-state  Potts model.}
\label{2DW3}
\end{figure}

We have studies the allowed region of $(\Delta_{\phi},\Delta_{n}')$ for CFTs with $S_3$ symmetry in two space-time dimensions. This result is presented in Figure \ref{2DW3}. 
We found that ``minimal models'' of $\mathcal{W}_3$ algebra, as classified in \cite{Fateev:1987vh}, saturate the unitarity bound. $\mathcal{W}_3$ algebra is an extension of the Virasoro algebra introduced by Zamolodchikov in \cite{Zamolodchikov:1985wn}. It contains the Virasoro algebra as a subalgebra. Besides the usual spin-2 operators $L_n$,  $\mathcal{W}_3$ algebra contains spin-3 operators $W_n$ which satisfies non-trivial commutation relations with $L_n$ and among themselves. Like for the Virasoro algebra, ``minimal models'' here means the fusion rules of the models consist of finite number of irreducible representations of $\mathcal{W}_3$.  It was shown in \cite{Fateev:1987vh}  that all these models have a global $Z_3$ symmetry, taking into account the complex conjugation of complex scalars one get the symmetric group $S_3$. The central charges of these models and the scaling dimensions of their $\mathcal{W}_3$ irreducible representations are given by,
\be
C_p = 2\big(1-\frac{12}{p(p-1)}\big)
\ee

\bea
 \Delta\bigg[\Phi \left(
\begin{array}{cc}
 n & m \\
n' & m' \\
\end{array}
\right)\bigg]&=&\frac{1}{12 p (p + 1)}\bigg(3 ((p + 1) (n + n') - p (m + m'))^2 \nn &&\quad\qquad\qquad+ ((p + 1) (n - n') - 
   p (m - m'))^2 - 12
\bigg),  
\eea
where $m,n,m',n'$ and $p$ are positive integers whose range are $n+n'\leq p-1$, $m+m'\leq p$ and $p\geq4$.
The horizontal and vertical axis in Figure \ref{2DW3} corresponds to operators with 
\be
\Delta_{\phi}=2 \times \Delta \bigg[\Phi\left(
\begin{array}{cc}
 1 & 2 \\
1 & 1 \\
\end{array}
\right)\bigg]=\frac{2 (p-3)}{3 (p+1)},\quad \text{and }\quad 
\Delta_{n}'=2 \times \Delta\bigg[\Phi\left(
\begin{array}{cc}
 1 & 3 \\
1 & 1 \\
\end{array}
\right)\bigg]=\frac{4 (2 p-3)}{3 (p+1)},
\ee
respectively. They satisfy 
\be
\Delta_{n}'=\frac{5}{2}\Delta_{\phi}+1,
\ee
which saturates the numerical bootstrap bound. It was discovered in \cite{Rychkov:2009ij} that minimal models of the Virasoro algebra also saturate the numerical bootstrap bound for CFTs with $Z_2$ symmetry. It is interesting to observe that $\mathcal{W}_3$ algebra also share the same feature. It would be interesting to extend this result to other $\mathcal{W}$-algebras.

\section{Discussion}
We have shown that there exist two series of conformal fixed points approach (de)coupled Ising model and Lee-Yang edge singularity respectively at the large $N$ limit. It would be interesting to understand whether it is possible to replace the large $N$ limit by other CFTs such as XY-model, Heinsberg model and etc. A naive guess is the following. The CFTs that approach Lee-Yang edge singularity has the symmetry group $S_N\otimes \mathbbm{1}$, while the CFTs that approach Ising model has $S_N\otimes Z_2$. It is therefore natural to consider scalar models with symmetry group $S_N\otimes G$, as a candidate for large $N$ CFTs that approach a CFT with symmetry group G. We leave this for future investigation.

In section \ref{SNboot}, we have shown that one could observe the fixed point $P_1$ in the numerical bootstrap curve, it would be interesting study its spectrum more carefully. The best way to do this is probably by first studying the possibility of isolating this fixed point using mixed correlator bootstrap, along the line of research in \cite{Kos:2014bka,Kos:2015mba,Li:2016wdp,Kos:2016ysd}. For the $N=4$ special case, a further comparison with experiment or Monte Carlo study would also be interesting.  It is also desirable to try to extract the $\mathcal{O}(1/N)$ corrections to the operator dimensions and compare them with our numerical bootstrap result. Since the $\mathcal{O}(1/N)$ effect receives contribution from all order in $\epsilon$-expansion, a proper resumption is necessary. What's more, it would be more interesting to investigate the possibility of performing a proper large-$N$ calculation like in O($N$) vector model (see \cite{Moshe:2003xn} for a review).

Before we close, let's think about the large $N$ (de)coupled CFTs in the context of AdS/CFT correspondence. As explained in section \ref{decoupleCFT}, the large $N$ spectrum of (de)coupled CFTs naturally fall into the categories of ``single trace operators'', ``double trace operators'' and so on. The replacement \eqref{changeboundary}, as famously pointed out by Witten \cite{Witten:1998qj}, corresponds to the change of boundary conditions for the dual AdS scalar, and does not change the dual  AdS mass as $M^2_{\text{AdS}}L^2=-\Delta(D-\Delta)$. The exact same phenomenon happens for O($N$) vector models. At the free theory limit, the scaling dimension of the first O($N$) singlet operator is given by $\Delta[{\sum_i\phi^i\phi^i}]=1$, while at the critical O($N$) point, its dimension is given by $D-1=2$ plus $\frac{1}{N}$ corrections. It is not yet clear what are the necessary and sufficient conditions for a CFT to have weakly coupled dual description in AdS \cite{Heemskerk:2009pn, Heemskerk:2010ty,ElShowk:2011ag,Fitzpatrick:2012cg,Hartman:2014oaa,Costa:2017twz}. As conjecture in \cite{Heemskerk:2009pn}, besides large $N$ factorization, any CFT with Einstein like local bulk dual description must also have a large gap for all single trace operators with spin higher than 2. This is clearly not the case for the large $N$ limit of decoupled CFTs. As shown in Section \ref{decoupleCFT}, the operators that could be interoperated as ``single trace'' operators are simply the S-channel operators of the component CFT, which clearly contains operators with arbitrary spin. If the dual theory indeed exist, it should be more similar to Vasiliev's higher spin theory \cite{Vasiliev:1990en,Vasiliev:1999ba}. However, since the CFT operators does not saturate the unitarity bound, higher spin symmetry is clearly broken in this case.

\section*{Acknowledgements}
We would like to thank Youjin Deng, Zhijin Li, Ziyang Meng and Yi Pang for helpful discussion and comments. The work of NS is supported by ITP-CAS. The numerical computations in this work are partially supported by HPC Cluster of SKLTP/ITPCAS.

\appendix 
\section{Renormalization of scalar field theory theory}\label{renormalization}
\subsection{3-Loop Renormalization of generic $\phi^4$ theory in $4-\epsilon$ Dimensions}\label{general3loop4D}
Suppose a group preserves a totally symmetric invariant tensor $d_{ijk}$, one can define the following constants $\{T_2,T_3,T_5,T_{71},T_{72}\}$ as \cite{Gracey:2015tta}
\bea\label{Tconstants}
&& d_{i_1 i_3 i_4}d_{i_2 i_3 i_4}=T_2 \delta_{i_1 i_2}\nn
&& d_{i i_1 i_2}d_{j i_1 i_3}d_{k i_2 i_3}=T_3 d_{ijk}\nn
&& d_{i i_1 i_2}d_{j i_3 i_4}d_{k i_5 i_6}d_{i_1 i_3 i_5}d_{i_2 i_4 i_6}=T_5 d_{ijk}\nn
&& d_{i i_1 i_2}d_{j i_3 i_4}d_{k i_5 i_6}d_{i_1 i_3 i_7}d_{i_2 i_5 i_8}d_{i_4 i_6 i_9}d_{i_7 i_8 i_9}=T_{71} d_{ijk}\nn
&& d_{i i_1 i_2}d_{j i_3 i_4}d_{k i_5 i_6}d_{i_1 i_3 i_7}d_{i_2 i_5 i_8}d_{i_4 i_8 i_9}d_{i_6 i_7 i_9}=T_{72} d_{ijk}\,.
\label{Tconstants}
\eea
Using the general formula summarised in \cite{Fei:2016sgs}, we could calculate the scaling dimensions of operators in $4-\epsilon$ up to two loop order. For a scalar field theory given by the Lagrangian \eqref{rescritedPotts}, we get the beta function 
\bea\label{betafunction}
\beta_1&=&-\epsilon g_1
+\frac{1}{C}\big[A_1 g_1^2+A_2 g_1 g_2+A_3 g_2^2+A_4 g_1^3 +A_5 g_1^2 g_2+A_6 g_1 g_2^2+ A_7 g_2^3\big]\nn
\beta_2&=&-\epsilon g_2+\frac{1}{C}\big[B_1 g_1^2+B_2 g_1 g_2+B_3 g_2^2+B_4 g_1^3 +B_5 g_1^2 g_2+B_6 g_1 g_2^2+ B_7 g_2^3\big]\nn
\eea
with the coefficient given by Table \ref{coefficients}.
\begin{table}[h]
\centering

\resizebox{\textwidth}{!}{

\begin{tabular}{|l|l|}
\hline
$A_1$ & $16 \pi ^2 n T_2^2+96 \pi ^2 n T_3 T_2+256 \pi ^2 n T_3^2+64 \pi ^2 n T_5-288 \pi ^2 T_2^2-192 \pi ^2 T_3 T_2+512 \pi ^2 T_3^2+128 \pi ^2 T_5$                      \\ \hline
$A_2$ & $192 \pi ^2 n T_2+384 \pi ^2 n T_3-384 \pi ^2 T_2+768 \pi ^2 T_3$                                                                                                   \\ \hline
$A_3$ & $0$                                                                                                                                                                 \\ \hline
$A_4$ & $n T_2^3+2 n T_3 T_2^2-20 n T_3^2 T_2-8 n T_5 T_2-64 n T_3^3-32 n T_3 T_5-32 n T_{71}+30 T_2^3+92 T_3 T_2^2+88 T_3^2 T_2+16 T_5 T_2-128 T_3^3-64 T_3 T_5-64 T_{71}$ \\ \hline
$A_5$ & $-32 n T_2^2-112 n T_3 T_2-192 n T_3^2-48 n T_5+256 T_2^2+64 T_3 T_2-384 T_3^2-96 T_5$                                                                              \\ \hline
$A_6$ & $-5 n^2 T_2-10 n^2 T_3-72 n T_2-184 n T_3+164 T_2-328 T_3$                                                                                                          \\ \hline
$A_7$ & $0$                                                                                                                                                                 \\ \hline
$B_1$ & $128 \pi ^2 T_2^3+320 \pi ^2 T_3 T_2^2-128 \pi ^2 T_3^2 T_2-128 \pi ^2 T_5 T_2$                                                                                     \\ \hline
$B_2$ & $64 \pi ^2 n T_2^2+128 \pi ^2 n T_3 T_2-128 \pi ^2 T_2^2+256 \pi ^2 T_3 T_2$                                                                                        \\ \hline
$B_3$ & $16 \pi ^2 n^2 T_2+32 \pi ^2 n^2 T_3+96 \pi ^2 n T_2+320 \pi ^2 n T_3-256 \pi ^2 T_2+512 \pi ^2 T_3$                                                                \\ \hline
$B_4$ & $-16 T_2^4-76 T_3 T_2^3-112 T_3^2 T_2^2+32 T_3 T_5 T_2+64 T_{71} T_2$                                                                                               \\ \hline
$B_5$ & $-5 n T_2^3-20 n T_3 T_2^2-20 n T_3^2 T_2-86 T_2^3-240 T_3 T_2^2+56 T_3^2 T_2+96 T_5 T_2$                                                                           \\ \hline
$B_6$ & $-44 n T_2^2-88 n T_3 T_2+88 T_2^2-176 T_3 T_2$                                                                                                                     \\ \hline
$B_7$ & $-9 n^2 T_2-18 n^2 T_3-24 n T_2-120 n T_3+84 T_2-168 T_3$                                                                                                           \\ \hline
$C$   & $256 \pi ^4 \left(n T_2+2 n T_3-2 T_2+4 T_3\right)$                                                                                                                 \\ \hline
\end{tabular}

}

\caption{Coefficients that appears in $\beta$ function}
\label{coefficients}
\end{table}

The anomalous dimensions are given by 
\bea\label{anadimension}
\gamma_{\phi}&=&\frac{g_2^2 (n+2)+g_1^2 T_2^2+2 g_1 T_2 \left(g_1 T_3+2 g_2\right)}{1024 \pi ^4},\nn
\gamma_{\phi^2\in S}&=&\frac{16 \pi ^2 \left(g_2 (n+2)+2 g_1 T_2\right)-3 \left(g_2^2 (n+2)+g_1^2 T_2^2+2 g_1 T_2 \left(g_1 T_3+2 g_2\right)\right)}{256 \pi ^4},\nn
\gamma_{\phi^2\in n}&=&\frac{ \left(g_1 T_2+2 g_1 T_3+2 g_2\right)}{16 \pi ^2}\nn
&&-\frac{g_2^2 (n+6)+g_1^2 \left(T_3 T_2+6 T_3^2+2 T_5\right)+8 g_2 g_1 \left(T_2+T_3\right)}{256 \pi ^4},\nn
\gamma_{\phi^2\in T'}&=&\frac{1}{256 \pi ^4 (n-2) (n+1)}\bigg(-g_2^2 ((n-2)) (n+1) (n+6)-4 g_1 g_2 \left(\left(n^2+n-6\right) T_2\right)
\nn
&&\qquad\qquad\qquad +32 \pi ^2 \left(g_1 (n-2) T_2-2 g_1 n T_3+g_2 (n-2) (n+1)\right)\nn
&& \qquad\qquad\qquad +g_1^2 \left(4 n \left(3 T_3^2+\text{T5}\right)+(6-7 n) T_2^2-4 (n-3) T_3 T_2\right) \bigg)\ldots.
\eea
Since the symmetric group $S_N$ also preserves a totally symmetric invariant tensor $d_{ijk}$, they fall into the type of models that could be calculated using the above formulas. Using the explicit construction of $d_{ijk}$ in \cite{Zia:1975ha}, it is easy to calculated the constants that appear in \eqref{Tconstants}, they are 
\bea\label{pottsT}
&&T_2=\frac{(n-1) (n+1)^2}{n^3},\nn
&&T_3=\frac{(n-2) (n+1)^2}{n^3},\nn
&&T_5=\frac{\left((n-2)^2+1\right) (n+1)^4}{n^6},\nn
&&T_{71}=\frac{(n+1)^6 \left((n+1)^3-9 (n+1)^2+29 (n+1)-32\right)}{n^9},\nn
&&T_{72}=\frac{(n-2) (n+1)^6 \left((n+1)^2-6 (n+1)+11\right)}{n^9}.
\eea
Plugging them into \eqref{betafunction}, solving $\beta_1=\beta_2=0$, we find the four fixed points in \eqref{fixedpointsPotts}. The free fixed is not renormalised. For other points, we could use \eqref{anadimension} to get the spectrum. For $\phi$, we have 
\bea\label{SDphi}
\Delta_{\phi}^{O(N)}&=&1-\frac{\epsilon }{2}+\frac{(n+2) \epsilon ^2}{4 (n+8)^2}+\ldots,\nn
\Delta_{\phi}^{P_1}&=&1-\frac{\epsilon}{2}+\frac{\left(n^2+8 n+7\right) \epsilon ^2}{108 (n+3)^2}+\ldots, \nn
\Delta_{\phi}^{P_2}&=&1-\frac{\epsilon}{2}+\frac{\left(n^4-9 n^3+31 n^2-45 n+22\right) \epsilon ^2}{108 \left(n^2-5 n+8\right)^2}+\ldots.
\eea
For quadratic operator in the S-channel, we have 
\bea\label{SDphi2S}
\Delta_{\phi^2\in S}^{O(N)}&=&2-\frac{6 \epsilon }{n+8}+\frac{(n+2) (13 n+44) \epsilon ^2}{2 (n+8)^3}+\ldots\nn
\Delta_{\phi^2\in S}^{P_1}&=&2-\frac{(n+7) \epsilon }{3 n+9}-\frac{\left(57 n^4-546 n^3+2016 n^2+7906 n+5159\right) \epsilon ^2}{486 (n-5) (n+3)^3}+\ldots\nn
\Delta_{\phi^2\in S}^{P_2}&=&2-\frac{2 \left(n^2-6 n+11\right) \epsilon }{3 \left(n^2-5 n+8\right)}+\nn
&&\frac{\epsilon ^2}{486 (n-5) \left(n^2-5 n+8\right)^3}\bigg(57 n^7-879 n^6+6174 n^5-26108 n^4\nn&&+69863 n^3-112629 n^2+96698 n-33176\bigg)+\ldots
\eea
For quadratic operator in the n-channel, we have 
\bea\label{SDphi2n}
\Delta_{\phi^2\in n}^{O(N)}&=&2+\left(\frac{2}{n+8}-1\right) \epsilon+\frac{\left(-n^2+18 n+88\right) \epsilon ^2}{2 (n+8)^3}+\ldots\nn
\Delta_{\phi^2\in n}^{P_1}&=&2-\frac{2 (n+4) \epsilon }{3 (n+3)}+\frac{\left(57 n^4+624 n^3-954 n^2-4832 n-3151\right) \epsilon ^2}{486 (n-5) (n+3)^3}+\ldots\nn
\Delta_{\phi^2\in n}^{P_2}&=&2-\frac{2 \left(n^2-5 n+9\right) \epsilon }{3 \left(n^2-5 n+8\right)}+\frac{\epsilon^2}{486 (n-5) \left(n^2-5 n+8\right)^3}\bigg(57 n^7-1137 n^6\nn
&&+9600 n^5-45914 n^4+ 135525 n^3-244965 n^2+247034 n-105384\bigg)\nn
&&+\ldots
\eea 
For quadratic operator in the T'-channel, we have 
\bea\label{SDphi2T}
\Delta_{\phi^2\in T'}^{O(N)}&&=2+\left(\frac{2}{n+8}-1\right) \epsilon+\frac{\left(-n^2+18 n+88\right) \epsilon ^2}{2 (n+8)^3}+\ldots\nn
\Delta_{\phi^2\in T'}^{P_1}&&=2+\left(\frac{2}{3 (n+3)}-1\right) \epsilon\nn
&&+\frac{\left(9 n^6-345 n^5+12 n^4-1614 n^3-2179 n^2+10311 n+10318\right) \epsilon ^2}{486 (n-5) (n-2) (n+1) (n+3)^3}+\ldots\nn
\Delta_{\phi^2\in T'}^{P_2}&&=2+\left(\frac{2}{3 \left(n^2-5 n+8\right)}-1\right) \epsilon \nn
&&+\frac{\epsilon ^2}{486 (n-5) (n-2) (n+1) \left(n^2-5 n+8\right)^3}\bigg( 9 n^9-180 n^8+1797 n^7-10116 n^6\nn
&&+33339 n^5-64960 n^4+66779 n^3-2712 n^2-79940 n+66352\bigg)+\ldots.
\eea 
The scaling dimension of quartic operator could be calculated using the eigenvalue of the matrix $\frac{\partial\beta_{i}}{\partial \lambda_j}$, for $P_1$, the final result turns out to be 
\bea\label{SDphi4SP1}
\Delta^{P_1}_{\phi^4\in S, \text{ 1st}}&=&4-\frac{2 (n+7) \epsilon }{3 (n+3)}\nn
&&+\frac{\left(-57 n^6-387 n^5+2808 n^4-13966 n^3-19345 n^2+53713 n+57106\right) \epsilon ^2}{243 (n+3)^3 \left(n^3-39 n+70\right)}\nn&&+\ldots\nn
\Delta^{P_1}_{\phi^4\in S, \text{ 2nd}}&=&4-\frac{\left(51 n^4+546 n^3+3060 n^2+5990 n+3409\right) \epsilon ^2}{81 (n+1) (n+3)^2 (n+7)}+\ldots
\eea
while for the point $P_2$
\bea\label{SDphi4SP2}
\Delta^{P_2}_{\phi^4\in S, \text{1st}}&=&4-\frac{4 \left(n^2-6 n+11\right) \epsilon }{3 \left(n^2-5 n+8\right)}\nn
&&+\frac{n-1}{243 (n-5) (n+1) \left(n^2-6 n+11\right) \left(n^2-5 n+8\right)^3}\bigg(57 n^9-1107 n^8\nn&&+9585 n^7-48407 n^6+154009 n^5-300181 n^4+280147 n^3+107591 n^2\nn&&-503846 n+333256\bigg)+\ldots\nn
\Delta^{P_2}_{\phi^4\in S, \text{ 2nd}}&=&4+\frac{\epsilon^2}{81 (n-2) \left(n^2-6 n+11\right) \left(n^2-5 n+8\right)^2} \bigg(78364 - 177712 n\nn&& + 175263 n^2 - 98431 n^3 + 34228 n^4 - 7398 n^5 + 
 921 n^6 - 51 n^7\bigg)+\ldots\nn
\eea
It is useful to record the renormalizaion of Ising model here for comparison 
\bea
\Delta_{\sigma}^{\text{Ising}}&=&1-\frac{\epsilon}{2}+\frac{\epsilon ^2}{108}+\ldots,\nn
\Delta_{\epsilon}^{\text{Ising}}&=&2-\frac{2 }{3}\epsilon+\frac{19}{162}\epsilon^2+\ldots,\nn
\Delta_{\epsilon'}^{\text{Ising}}&=&4-\frac{17}{27} \epsilon ^2+\ldots.
\eea
This result is taken from, for example, \cite{Kleinert:2001ax}.
\subsection{3-Loop Renormalization of generic $\phi^3$ theory in $6-2\epsilon$ Dimensions}\label{general3loop6D} 
Three loop renormalization of generic $\phi^3$ theory in $D=6-2\epsilon$ was studied by \cite{deAlcantaraBonfim:1980pe,deAlcantaraBonfim:1981sy}. Four loop result was obtained more recently in \cite{Gracey:2015tta,Gracey:2017uzx}, where they have also studied the renormalization of the Potts model and the Lee-Yang edge Singularity. The authors did not present the result for $N$-state Potts model with generic $N$, but rather focus on the $N\rightarrow 1$ limit to study percolation problem. For the reader's convenience, we will record the generic $N$ result here. 
Plug \eqref{pottsT} into the formulas in \cite{Gracey:2015tta}, one could easily get
\bea\label{phi3all}
\Delta_{\phi}&=&2-\frac{2 (5 n-11) \epsilon }{3 (3 n-7)}-\frac{2 (n-1) \left(43 n^2-171 n+206\right) \epsilon ^2}{27 (3 n-7)^3}\nn
&&+\frac{(n-1) \epsilon ^3}{243 (3 n-7)^5} \bigg(15552 n^4 \zeta (3)-8375 n^4-129600 n^3 \zeta (3)+68025 n^3\nn&&+466560 n^2 \zeta (3)-210179 n^2-829440 n \zeta (3)+300903 n\nn&&+580608 \zeta (3)-187238\bigg),\nn
\Delta_{\phi^2\in S}&=&4-\frac{8 (n-4) \epsilon }{3 (3 n-7)}+\frac{2 \left(43 n^3-247 n^2+857 n-653\right) \epsilon ^2}{27 (3 n-7)^3}\nn
&&+\frac{1}{243 (3 n-7)^5}\epsilon^3\bigg(-15552 n^5 \zeta (3)+8375 n^5+28512 n^4 \zeta (3)-66665 n^4\nn&&-207360 n^3 \zeta (3)+163514 n^3+1467072 n^2 \zeta (3)\nn
&&-224126 n^2-2887488 n \zeta (3)+450911 n+1614816 \zeta (3)-332009\bigg),\nn
\Delta_{\phi^3\in S}&=&6+\frac{2 \left(-125 n^2+544 n-671\right) \epsilon ^2}{9 (3 n-7)^2}+\frac{\epsilon^3}{81 (3 n-7)^4}\bigg(38880 n^4 \zeta (3)+36755 n^4\nn
&&-316224 n^3 \zeta (3)-319602 n^3+1187136 n^2 \zeta (3)+1123920 n^2-2265408 n \zeta (3)\nn&&-1831190 n+1687392 \zeta (3)+1097253\bigg).
\eea
We also record here the renormalization for Lee-Yang edge singularity for comparison, setting
\be
T_2=T_3=T_5=T_{71}=T_{72}=1,
\ee
one get 
\bea
\Delta_{\phi}&=&2-\frac{10\epsilon }{9}-\frac{86 \epsilon ^2}{729}+\left(\frac{64 \zeta (3)}{243}-\frac{8375}{59049}\right) \epsilon ^3+\mathcal{O}(\epsilon^4),\nn
\Delta_{\phi^3}&=&6-\frac{250 \epsilon ^2}{81}+\left(\frac{160 \zeta (3)}{27}+\frac{36755}{6561}\right) \epsilon ^3+\mathcal{O}(\epsilon^4).
\eea
It is not necessary to present the dimension of $\Delta_{\phi^2}$, since it is the conformal descendent of $\phi$. As a result of equation of motion $\Box \phi \sim \phi^2$, its dimension is fixed to be $\Delta_{\phi^2}=\Delta_{\phi}+2$.

\section{Bootstrap with $S_N$ symmetry}\label{Snsection}
Using the ``villeins'' $e^{\alpha}_i$, beside $d_{ijk}$ defined in \eqref{dtensor}, one could also define the following invariant tensor carrying four indices
\bea
Q_{ijkl}&=&\sum_{\alpha}e_i^{\alpha}e_j^{\alpha}e_k^{\alpha}e_l^{\alpha},
\eea
They satisfy   
\be
d_{ijm}d_{klm}=\frac{n+1}{n}Q_{ijkl}-\frac{(n+1)^2}{n^3}\delta_{ij}\delta_{kl}
\ee
and 
\be
d_{ikl}d_{jkl}=\frac{(n-1)(n+1)^2}{n^3}\delta_{ij}.
\ee
The production of two $n$-dimensional representation can be decomposed as,
\be
\text{n}\otimes \text{n} \rightarrow \text{S} \oplus \text{A} \oplus \text{n}\oplus \text{T}'.  \nonumber
\ee
Compare with the production rule for rotational group $O(n)$, $\text{n}\otimes \text{n} \rightarrow \text{S} \oplus \text{A} \oplus \text{T}$, the $T$ representation of $O(n)$ group is further decomposed into n$\oplus$ T$'$, due to the existence of $d_{ijk}$.
One could also defines the following linear independent invariant tensors 
\bea
P^{(1)}_{ijkl}&=&\frac{1}{n}\delta_{ij}\delta_{kl},\nonumber\\
P^{(n)}_{ijkl}&=&\frac{n^3}{(n-1)(n+1)^2}d_{ijm}d_{klm},\nonumber\\
P^{(T')}_{ijkl}&=&\frac{1}{2}\delta_{il}\delta_{jk}+\frac{1}{2}\delta_{ik}\delta_{jl}-\frac{1}{n}\delta_{ij}\delta_{kl}-\frac{n^3}{(n-1)(n+1)^2}d_{ijm}d_{klm},\nonumber\\
P^{(A)}_{ijkl}&=&\frac{1}{2}\delta_{il}\delta_{jk}-\frac{1}{2}\delta_{ik}\delta_{jl}.
\eea
Suppose $v_1^i$ and $v_2^i$ are two vectors carrying indices in n-dimensional representation of $S_N$, the tensors 
\be
P^{(I)}_{ijkl}v_1^kv_2^l
\ee
transforms in irreducible representation ``$I$'' of $S_N$ group. It could be checked that these projectors satisfies the following relations
\bea
&&P^{(I)}_{ijmn}P^{(I)}_{nmkl}=P^{(I)}_{ijkl},\nn
&&P^{(I)}_{ijkl}\delta_{il}\delta_{jk}=\text{dim}_I.
\eea
where $\text{dim}_I$ stands for the dimension of representation $I$. 

A four point function in CFTs with $S_N$ global symmetry can be written as
\bea\label{F4equations}
\langle
\contraction{}{\phi_i(x_1)}{}{\phi_j(x_2)}
\phi_i(x_1)\phi_j(x_2)
\contraction{}{\phi_k(x_3)}{}{\phi_l(x_4)}
\phi_k(x_3)\phi_{l}(x_4)
\rangle=&& \frac{1}{x_{12}^{2\Delta_{\phi}}x_{34}^{2\Delta_{\phi}}}\sum_I {P}^{(I)}_{ijkl}\left(\sum_{{\cal O}\in I}\lambda_{\cal O}^2 g_{\Delta_{\cal O},l_{\cal O}}(u,v)\right)\nn
\text{where}\quad  I\in\{{1}^+,{n}^+,{T'}^+,{A}^-\}.&&
\eea
Here $I^{\pm}$ denotes operators with even(odd) spin and transforms in  irreducible representation ``$I$'' of $S_N$. See \cite{Rattazzi:2010yc} for the reason behind the spin choice. $g_{\Delta_{\cal O},l_{\cal O}}(u,v)$ is the conformal block which encodes all the kinematics of conformal field theories, which is universal for any CFTs. The dynamical information which are specific to each CFT, on the other hand, are widely believed to be encoded in the OPE coefficients and the spectrum. An analytical expression for conformal block in even dimension was calculated in \cite{Dolan:2000ut,Dolan:2003hv}. Operator product expansion are convergent for conformal field theories, and four point functions should not depend on how OPE is preformed, so
\be\label{crossingsym}
\langle
\contraction{}{\phi_i(x_1)}{}{\phi_j(x_2)}
\phi_i(x_1)\phi_j(x_2)
\contraction{}{\phi_k(x_3)}{}{\phi_{l}(x_4)}
\phi_k(x_3)\phi_{l}(x_4)
\rangle=\langle
\contraction{}{\phi_i(x_1)}{\phi_j(x_2)\phi_k(x_3)}{\phi_{l}(x_4)}
\contraction[2ex]{\phi_i(x_1)}{\phi_i(x_2)}{}{\phi_{l}(x_3)}
\phi_i(x_1)\phi_j(x_2)\phi_k(x_3)\phi_{l}(x_4)
\rangle.
\ee
From this equality we get the following crossing equations 
\bea\label{crossingeqn2}
\sum_{I}\sum_{{ O}\in I}\lambda_{\phi \phi O}^2\vec{V}^{(I)}_{\Delta_{O},l_{\cal O}}(u,v)=0\,,\quad \text{with}\quad  I\in\{{1}^+,{n}^+,{T'}^+,{A}^-\}\,,
\eea
where 
\be
\vec{V}^{(1^+)}_{\Delta_{O},l_{\cal O}}(u,v)=
\left(
\begin{array}{c}
 0 \\
 0 \\
 \frac{F}{n} \\
 -\frac{H}{n} \\
\end{array}
\right),
\vec{V}^{(n^+)}_{\Delta_{O},l_{\cal O}}(u,v)=
\left(
\begin{array}{c}
 F \\
 0 \\
 \frac{F}{1-n} \\
 \frac{H}{n-1} \\
\end{array}
\right),
\ee
\be
\vec{V}^{(T'^+)}_{\Delta_{O},l_{\cal O}}(u,v)=
\left(
\begin{array}{c}
 -F \\
 \frac{F}{2} \\
 \frac{F \left(n^2-n+2\right)}{2 (n-1) n} \\
 \frac{H \left(n^2-n-2\right)}{2 (n-1) n} \\
\end{array}
\right),
\vec{V}^{(A^-)}_{\Delta_{O},l_{\cal O}}(u,v)=
\left(
\begin{array}{c}
 0 \\
 -\frac{F}{2} \\
 \frac{F}{2} \\
 \frac{H}{2} \\
\end{array}
\right).
\ee
Here $F$ and $H$ are short for $F_{\Delta,l}$ and $H_{\Delta,l}$, defined by
\bea
F_{\Delta,l}&=&\frac{v^{\Delta_{\phi}}G_{\Delta,l}(u,v)-u^{\Delta_{\phi}}G_{\Delta,l}(v,u)}{u^{\Delta_{\phi}}-v^{\Delta_{\phi}}},\nn
H_{\Delta,l}&=&\frac{v^{\Delta_{\phi}}G_{\Delta,l}(u,v)+ u^{\Delta_{\phi}}G_{\Delta,l}(v,u)}{u^{\Delta_{\phi}}+ v^{\Delta_{\phi}}}.
\eea
The logic for numerical bootstrap is to look for a linear functional $\alpha$ such that 
\begin{alignat}{4}\label{linearfunctional}
&\alpha(\vec{V}^{({1}^+)}_{0,0})=1\,,   &\qquad\qquad &\nn
&\alpha(\vec{V}^{(I)}_{\Delta,0})\geq0\,, & &\text{for }\Delta\geq \frac{D-2}{2} \,,\quad  I\in\{1^+,n^+,T'^+\}\,,\nn
&\alpha(\vec{V}^{(n^+)}_{\Delta,0})\geq0\,, & &\text{for }\Delta\geq \Delta_{n}\,,\nn
&\alpha(\vec{V}^{(I)}_{\Delta,l})\geq 0\,,  & &\text{for }\Delta\geq l+D-2\,, (l=2,4,6,8,10\dots)\,\, \text{ and }  I\in\{1^+,n^+,T'^+\},\nn 
&\alpha(\vec{V}^{(A^-)}_{\Delta,l})\geq 0\,, & & \text{for }\Delta\geq l+D-2 \,,(l=1,3,5,7,9\dots)\,.
\end{alignat}
This realise the conditions imposed on the operator spectrum in section \ref{SNboot} to study conformal field theories with $S_N\otimes Z_2$ symmetry. If such a functional could be found, then there is no way for \eqref{crossingeqn2} to be satisfied with all the $\lambda_{\mathcal O}^2$'s being positive. Therefore we conclude that a unitary CFT with $S_N\otimes Z_2$ symmetry and $\Delta_{\phi}$ must have at least one scalar operators whose dimension is less than $\Delta_{n}$. For readers interested in the implement of numerical bootstrap, we refer them to \cite{Simmons-Duffin:2015qma} and reference therein. The numerical computations in this work are performed using the SDPB package \cite{Simmons-Duffin:2015qma}. For the approximation of the conformal blocks, we partially used the code from JuliBoot \cite{Paulos:2014vya}.

Before proceeding, let's recall the dimensions of each representations to be, 
\be
dim_S=1, \quad dim_n=n,\quad dim_A=\frac{n(n-1)}{2},\quad dim_{T'}=\frac{n(n+1)}{2}-1-n.
\ee
For $n=2$, hence $S_3$ group, $dim_{T'}=0$, one can check that $P^{(T')}_{ijkl}=0$, and
\bea\label{O2projectors}
P^{(S)}_{ijkl}&=&\frac{1}{2}\delta_{ij}\delta_{kl},\nonumber\\
P^{(n)}_{ijkl}&=&\frac{1}{2}\delta_{il}\delta_{jk}+\frac{1}{2}\delta_{ik}\delta_{jl}-\frac{1}{2}\delta_{ij}\delta_{kl}=\frac{8}{9}d_{ijm}d_{klm},\nonumber\\
P^{(A)}_{ijkl}&=&\frac{1}{2}\delta_{il}\delta_{jk}-\frac{1}{2}\delta_{ik}\delta_{jl}.
\eea
which are the same projectors as for SO(2) group. Using these projectors, we could derived the following crossing equations
\bea\label{crossingeqn2}
\sum_{I}\sum_{{ O}\in I}\lambda_{\phi \phi O}^2\vec{V}^{(I)}_{\Delta_{O},l_{\cal O}}(u,v)=0\,,\quad \text{with}\quad  I\in\{{1}^+,{n}^+,{A}^-\}\,,
\eea
with 
\be
\vec{V}^{(1^+)}_{\Delta_{O},l_{\cal O}}(u,v)=\left(
\begin{array}{c}
 0 \\
 F \\
 H \\
\end{array}
\right),\quad
\vec{V}^{(n^+)}_{\Delta_{O},l_{\cal O}}(u,v)=\left(
\begin{array}{c}
 F \\
 0 \\
 -2 H \\
\end{array}
\right),\quad
\vec{V}^{(A^-)}_{\Delta_{O},l_{\cal O}}(u,v)=
\left(
\begin{array}{c}
 -F \\
 F \\
 -H \\
\end{array}
\right)
\ee
These are exactly the same crossing equations that were used for bootstraping O(2) invariant CFTs in \cite{Kos:2013tga}. However, when studying conformal field theories with $S_3$ symmetry, since $d_{ijm}$ is an invariant tensor of $S_3$ group (which is not invariant under SO(2) group), scalars $\phi^i$ would appear in its own OPE, $\phi^i \times \phi^j\sim d_{ijk} \phi^k$. We need to search for a linear functional $\alpha$ satisfying \eqref{linearfunctional} plus one extra condition 
\be
\alpha(\vec{V}^{(n^+)}_{\Delta_{\phi},0})\geq 0.
\ee
This is the numerical bootstrap program used in section \ref{W3result}.


\end{document}